\documentclass{article}
\usepackage{smc2026}
\usepackage[caption=false, font=footnotesize]{subfig}% Modern replacement for subfigure package
\usepackage{paralist}% extended list environments
\usepackage[figure,table]{hypcap}% hyperref companion

\usepackage[english]{babel}
\usepackage[whole]{bxcjkjatype}
\usepackage{adjustbox}
\usepackage{algorithm}
\usepackage{algpseudocode}
\usepackage{stfloats}

\def\papertitle{Precise and Simple Audio-to-Score Alignment}

\author[1, 2]{\mbox{\firstname{Silvan}\middlename{David}\lastname{Peter}}}
\author[1]{\mbox{\firstname{Patricia}\lastname{Hu}\originalname{胡紫漪}
}}
\author[1, 2]{\mbox{\firstname{Gerhard}\lastname{Widmer}}}%orcid{0000-0001-8437-4517}}}

\affil[1]{\institution{Institute of Computational Perception, Johannes Kepler University}\city{Linz}\country{Austria}\affiliationtype{University}}
\affil[2]{\institution{LIT AI Lab, Linz Institute of Technology}\city{Linz}\country{Austria}\affiliationtype{Music}}

\completesetup

\title{\papertitle}
\begin{document}
	\capstartfalse
	\maketitle
	\capstarttrue

\section{Introduction}\label{sec:introduction}
\begin{figure*}[!b]
    \centering
    {\includegraphics[width=2\columnwidth]{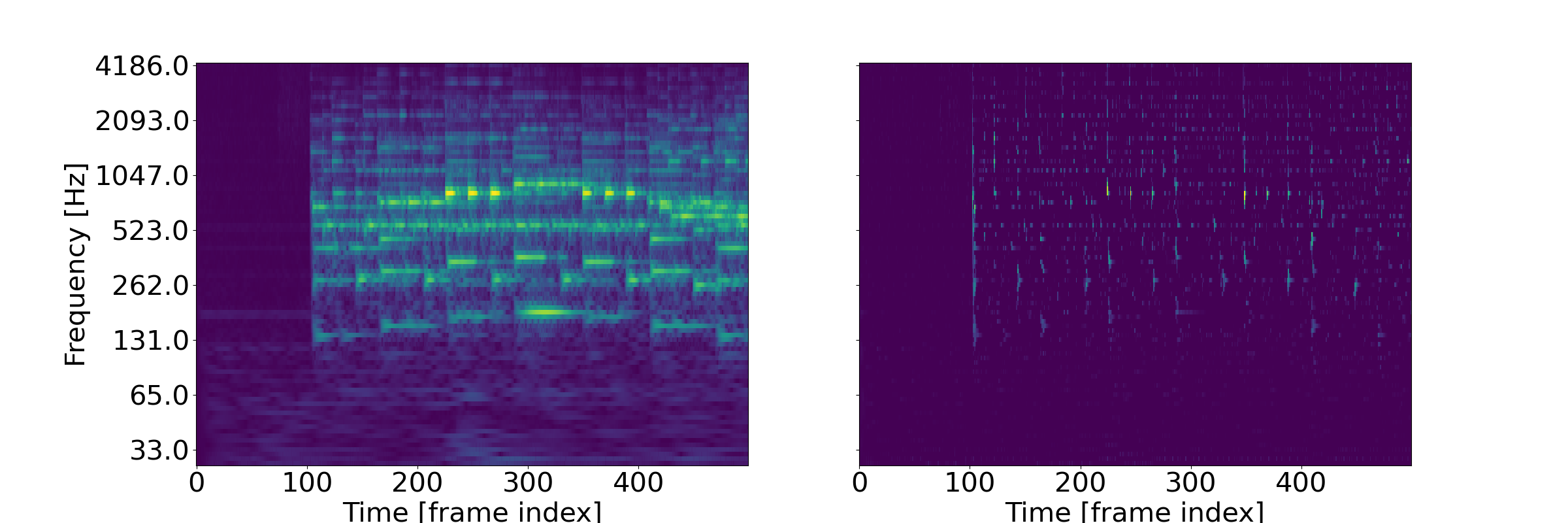}} 
    \caption{Spectral (left) and onset (right) activation features on the first ten seconds of a piano recording. There are 88 frequency bins (rows) logarithmically spaced and centered on the piano key frequencies. The temporal frame rate is 50 Hz.}
    \label{fig:features}
\end{figure*}

Audio-to-score alignment is a long-standing challenge in music information retrieval and arguably the most widely applicable alignment task for music research. Alignment algorithms match two versions of a piece of music, and for this to work these versions need to be in comparable formats. Audio-to-audio alignment matches audio features; when matching audio files to scores, they must either synthesize the score or derive audio-like features by means of piano rolls or similar feature sequences~\cite{kwon2017audio,murgul2025fine,zeitler2024robust}. Symbolic alignment, by contrast, matches symbolically encoded notes; in an audio-to-score scenario these would be obtained by a transcription of the audio file~\cite{simonetta2021audio,peter25transalign}. 
In this article, we present an algorithm that bridges audio-like and symbol-level features directly. Sequential audio features encoding onset and spectral activation (see Figure \ref{fig:features}) are matched to score positions by a bespoke dynamic programming-based matching algorithm derived from symbolic alignment methods. The resulting method is both precise - surpassing widely used audio-to-audio approaches based on synthesized scores -, and remains flexible in its digital signal processing components, i.e., the method is adaptable to diverse timbral characteristics without requiring a separate transcription model. Furthermore it inherits some of the symbolic alignment runtime advantages with an algorithmic complexity that is at worst linear in the length of the (typically short) symbolic score and (typically long) audio feature sequence. In the following sections, we provide a detailed algorithm description and evaluate its alignment quality on a large-scale dataset of solo piano recordings.

\begin{algorithm*}[t]
\caption{Score-to-Audio Alignment via Dynamic Programming}
% \centering
  % \adjustbox{max width=2\columnwidth}{
\begin{algorithmic}[1]

\Require 
$\texttt{score\_onset\_times}[i]$ \Comment score times in beats
\Require 
$\texttt{pitch\_sets}[i]$ \Comment notated pitches at score onset
\Require 
$\texttt{onsets}[p,t]$ \Comment onset activation for pitch $p$ at frame $t$
\Require 
$\texttt{spec}[p,t]$ \Comment spectral presence for pitch $p$ at frame $t$
\Require 
$\texttt{stretch\_limits}, \texttt{cost\_weights}$

\State $D[i,j] \gets \infty$ \Comment accumulated cost
\State $B[i,j] \gets -1$ \Comment backpointer
\State $BP[i,j] \gets bp_{\mathrm{init}}$ \Comment beat period estimate
\State $D[0,0] \gets 0$

\For{$i = 0$ to $M-1$}  \Comment loop over score onsets
    \For{$j = 0$ to $N-1$}  \Comment loop over audio frames

        \If{$D[i,j] = \infty$}
            \State \textbf{continue}
        \EndIf

        \State $bp \gets BP[i,j]$ \Comment get local beat period estimate
        \State $\Delta_{\mathrm{score}} \gets 
        \texttt{score\_onset\_times}[i+1]
        - \texttt{score\_onset\_times}[i]$ 

        \State $\texttt{candidate\_frames} \gets 
        \texttt{compute\_frame\_window}(j, bp, \Delta_{\mathrm{score}}, \texttt{stretch\_limits})$ 

        \ForAll{$p \in \texttt{pitch\_sets}[i+1]$}

            \ForAll{$j' \in \texttt{candidate\_frames}$}

                \State $\texttt{stretch\_term} \gets 
                \texttt{stretch\_cost}(j' - j, bp, \Delta_{\mathrm{score}})$

                \State $\texttt{onset\_term} \gets 
                \texttt{onsets}[p,j']$

                \State $\texttt{spec\_term} \gets 
                \min_k \bigl(\texttt{spec}[p,j'+k]\bigr)$

                \State $\texttt{transition\_cost} \gets D[i,j]$
                \Statex \hspace{\algorithmicindent}
                $+ \; w_{\mathrm{onset}} \cdot \texttt{onset\_term}$
                \Statex \hspace{\algorithmicindent}
                $+ \; w_{\mathrm{stretch}} \cdot \texttt{stretch\_term}$
                \Statex \hspace{\algorithmicindent}
                $+ \; w_{\mathrm{spec}} \cdot \texttt{spec\_term}$

                \If{$\texttt{transition\_cost} < D[i+1,j']$}

                    \State $D[i+1,j'] \gets \texttt{transition\_cost}$

                    \State $B[i+1,j'] \gets j$
                    \Comment backtracking pointer

                    \State $BP[i+1,j'] \gets 
                    \texttt{update\_beat\_period}(j' - j,\Delta_{\mathrm{score}},bp)$

                \EndIf

            \EndFor
        \EndFor
    \EndFor

    \State $D[i+1,\texttt{mask\_cost\_above\_reset\_threshold}] \gets \infty$

\EndFor

\State $\texttt{alignment} \gets \texttt{backtrack\_through\_pointers}(B)$
\State \Return $\texttt{alignment}$

\end{algorithmic}\label{code:0}
% }
\end{algorithm*}

\section{Alignment Method}

\subsection{Signal Processing}
The audio signal is processed into two feature sequences, one for onset (time) information, the other for spectral (pitch) information. As a first step, the stereo signal is summed to mono and then sent through an IIR filterbank of second-order Butterworth filters. The filterbank consists of 88 filters centered at the key frequencies of an equally tempered 440 Hz Chamber pitch piano. The passband limits are set to the quarter tone middle points between adjacent pitch frequencies. The default filterbank is set up for repertoire of this temperament, tuning, and register. Different setups are possible if the musical material is known to differ. The 88 filtered signals are aggregated by window-wise maximal values with window width and hop size being set for 50 Hz frame sequences and stacked as a spectrogram. 
The onset features are derived by a superflux algorithm\cite{bock} from the framed signal: a maximum filtered frame (across three vertically adjacent frequency bins) is subtracted from the subsequent frame and halfwave rectified. The resulting feature is bin-wise normalised to one. The durational feature is directly given by the original filtered and framed signal, again normalized to one for each frequency bin. While this procedure results in usable pitch-wise activation features for piano music, onset activation, normalization, and framerate can be adapted to suit the needs of different types of audio.
The score representation consists of a list of score chords (for simplicity just coinciding notes irrespective of durations or voices) given by their onset beat position and the MIDI pitches of the notes encoded at this position.

\subsection{Dynamic Programming}
The alignment algorithm treats the pitch-wise onset and spectral activations as a proto-transcription and relates it to score information. For any position in the score and each pitch expected to be played at it, all positions in a given temporal window of the framed signal are given a cost related to the best fitting onset position with subsequent spectral activation. In typical dynamic programming fashion it starts at the beginning of both sequences and works through all score and signal window positions while keeping track of previously best aligned subsequences. Pseudocode~\ref{code:0} outlines the algorithm structure.

\begin{table*}[t]]
\centering
  \adjustbox{max width=1.9\columnwidth}{
\begin{tabular}{lrrrrrr}
\textbf{Method} &
\textbf{Mean (ms)} &
\textbf{Median (ms)} &
\textbf{$< 50$ ms (\%)} &
\textbf{$< 100$ ms (\%)} &
\textbf{$< 200$ ms (\%)} &
\textbf{$< 500$ ms (\%)} \\ \hline
Audio-to-Audio         & 135 & 49 & 53.2 & 74.4 & 87.7 & 91.7 \\
Audio-to-Score (ours) & 86  & 21 & 83.7 & 91.7 & 95.2 & 97.9 \\
MIDI-to-Score         & 6   & 0  & 98.1 & 98.5 & 99.2 & 99.7 \\ \hline
\end{tabular}}
 \caption{Alignment results across a baseline audio-to-audio method using a synthesized score and onset as well as chroma features (first row), our proposed mixed audio-symbolic method (second row), and a MIDI-based symbolic alignment method (third row). Mean and median values are given in milliseconds, lower values are better. The remaining columns show the percentage of absolute errors below given thresholds, higher values are better. All values are averaged performance-wise across the dataset. 
 }
 \label{tab:results}
\end{table*}

The cost function relating score positions to onset times in the spectrogram combines three components: an onset term (strong onset activation), a spectral term (continued spectral energy for several frames), and a stretch term (favoring low time warping or tempo variation). The latter is influenced by a path-wise beat period estimate that is continuously updated, associating lower costs to less locally variable tempo estimates.  The algorithm affords several tuning parameters like the stretch limits and the weights to sum the cost from its stretch, onset, and spectral components. Each cost component is normalized to fall between zero and one.

\section{Evaluation}
We evaluate our algorithm on over 300 piano performances from the (n)ASAP Dataset~\cite{Peter-2023}. We compare it to an audio-to-audio alignment baseline which uses Dynamic Time Warping on both onset-related and spectral features. The implementation is given by the synctoolbox library~\cite{muller2021sync}. Audio-to-audio alignment based on a mix of features and synthesized audio is a common baseline for audio-to-score alignment. When high-quality transcriptions are available, symbolic alignment becomes a more precise baseline. To give an estimated upper bound for the quality of transcription-based symbolic alignment, we assess a MIDI-to-score alignment method (“DualDTWMatcher”) from the parangonar library~\cite{Peter-2023} using the recorded MIDI performances of the (n)ASAP Dataset as proxies for perfect transcriptions. Table~\ref{tab:results} shows the results in terms of mean and median errors as well as percentages of errors below four different thresholds.

Our method surpasses the baseline audio-to-audio method at all levels of precision, yet unsurprisingly falls short of the precision of a symbolic alignment model. Notably, several alignments were excluded from the audio-to-audio version where an obviously spurious alignment was computed, while both our method and the MIDI-to-score alignment worked robustly across the entire dataset. 
There is a runtime to precision tradeoff in the setting of the window size, frame rate, and threshold for cost rest. The higher these values, the more precise the alignment becomes, and the longer it takes to compute it. The values shown stem from a parameter set on the precise yet slow side (no threshold, medium window, 50 Hz). However, we did not optimize the parameter settings.

\section{Conclusion}

We introduce an audio-to-score algorithm which uses both onset and spectral audio features in a note-based matching procedure typically found in symbolic alignment. Our method leverages dynamic beat period estimates and score-informed pitch-wise onset and spectral processing to produce highly precise alignments. It relies on standard digital signal processing and dynamic programming techniques without the need for external processing through transcription, neural features, or synthesis. We hope that our implementation provides a simple and directly accessible tool for the community.

Our implementation is available online: 
\url{https://github.com/sildater/parangonar}

\begin{acknowledgments}
This research acknowledges support by the Linz Institute of Technology Artificial Intelligence Lab and the by the European Research Council (ERC), under the European Union's Horizon 2020 research and innovation programme, grant agreement No.~101019375 \textit{Whither Music?}.
\end{acknowledgments}

\bibliography{refs}
	
\end{document}